# Multi Site Coordination using a Multi-Agent System

Monteiro Thibaud[(1)], Roy Daniel[(2)], Anciaux Didier[(1)]

INRIA – MACSI project, LGIPM – SdP team
[(1)]Université Paul Verlaine, Île du Saulcy, F-57045 Metz CEDEX 01 – FRANCE
[(2)]École Nationale d'Ingénieurs de Metz, Île du Saulcy, F-57045 Metz CEDEX 01 – FRANCE
Email: {roy, anciaux, monteiro}@agip.sciences.univ-metz.fr

**Abstract**

A new approach of coordination of decisions in a multi site system is proposed. It is based this approach on a multi-agent concept and on the principle of distributed network of enterprises. For this purpose, each enterprise is defined as autonomous and performs simultaneously at the local and global levels.

The basic component of our approach is a so-called Virtual Enterprise Node (VEN), where the enterprise network is represented as a set of tiers (like in a product breakdown structure). Within the network, each partner constitutes a VEN, which is in contact with several customers and suppliers. Exchanges between the VENs ensure the autonomy of decision, and guarantiee the consistency of information and material flows. Only two complementary VEN agents are necessary: one for external interactions, the Negotiator Agent (NA) and one for the planning of internal decisions, the Planner Agent (PA).

If supply problems occur in the network, two other agents are defined: the Tier Negotiator Agent (TNA) working at the tier level only and the Supply Chain Mediator Agent (SCMA) working at the level of the enterprise network. These two agents are only active when the perturbation occurs. Otherwise, the VENs process the flow of information alone.

With this new approach, managing enterprise network becomes much more transparent and looks like managing a simple enterprise in the network. The use of a Multi-Agent System (MAS) allows physical distribution of the decisional system, and procures a heterarchical organization structure with a decentralized control that guaranties the autonomy of each entity and the flexibility of the network.

## 1. Introduction

Product design, manufacture, conditioning, or the combination of the three do not result from isolated and autarkical companies but from increasingly complex corporative networks. Such networks can take various forms and can be as complex as clusters of several "Industrial Architectures".

It is necessary to develop rigorous methods to improve the performance of such complex architecture development and the efficiency of these enterprises. In a complex network, enterprise performance is sensitive to the relationships and behaviors of participating companies, a dimension which does not exist for individual companies. Being a function of product quality and costs (including the societal cost) of the products, it is also defined in terms of time to design or to manufacture, and depends mainly on the information and the material flows.

To improve the reactivity and the costs of a company involved in such a complex architecture,

it is necessary to consider subcontracting and the relationships with partners. The "make or buy" decisions involved in this architecture occur in various time frames. A long term decision or strategic decision, corresponds to the whole set of external and internal nodes of production, distribution and supply. A medium term decision or tactic decision characterizes the contracts (quantities of products to deliver, delays, costs, penalties…) that the company is likely to have with its internal and/or external providers in order to carry out a production program that accomplishes the best balance between cost and delay. Finally, a short term decision or operational decision, considers for instance subcontractors, simply for overloading capacity in order to absorb a transitory fluctuation of demand. The paper only focuses on operational decisions.

There is a need to better coordinate actions in complex cooperative network, especially among the partners (Altersohn, 1992; Rota, 1998; Kjenstad, 1998). Recent research shows a growing interest in studying cooperation relationships among multiple actors of industrial architectures (Axelrod, 1992; Rapoport, 1987; Ferrarini, 2001; Monteiro and Ladet, 2001b). It appears from these studies that cooperation has two different forms. Cooperation can be a form of collaboration between partners in which each has equivalent decisional capacity and acts with others towards a common objective, such as in the co-design in the automotive sector (Womack *et al.*, 1992) or aeronautical sector (Cauvin *et al.*, 2003). Cooperation can also be a form of coordination and a synchronization of operations carried out by independent actors (Malone, and Crowston, 1994; Monteiro and Ladet, 2001a). In this case, each partner has a limited decision power that corresponds to its action field (Camalot, *et al.*, 1997; Camalot, 2000; Huguet, 1994).

Multi site resources are considered in this paper to study the planning management of complex enterprise networks. A performance criterion valid locally as well as globally is used, and a "win-win" policy, based on costs is used, that controls the network (OUZIZI, *et al.*, 2003).

In addition, the distribution of the enterprise network cannot be managed only by one and unique data-processing and data base application. The main reason is that the exchange of information and the behaviors, wich are specific to operations of the network members, are so complex that they require computing paradigms which need to be decentralized and shared. Consequently, it seems that a multi-agent architecture can best meet this need (Ferber, 1995; Patriti, *et al.*, 1997).

Therefore, a new approach is proposed. It is based on a cooperative multi-agent architecture in order to represent and manage the operational decisions made in complex enterprise networks.

First, relevant research work on the coordination of supply chains is reviewed. Then, a simple case for water tab production is described and used to develop the approach by presenting the model architecture and the agent specifications.

## 2. Literature review on coordination of the supply chains and enterprise networks

Nowadays, there is a large and growing number of research efforts on coordination and management of supply chains.

The first area concerns operational research models. The aim of these models is to coordinate distributed planning with a collaborative approach (Dudek and Stadler, 2005; Schneeweiss and Zimmer, 2004). Each partner, supplier or producer, is described in the mathematical model by an objective function. The models are combined to coordinate activities, but do not consider information flows and decisional processes.

The second area of research concerns collaborative architectures, mostly viewed as a centralized architecture. A large literature review on this area can be found in the paper of Stadtler (2005). Some research papers deal with distributed architectures. Verwijmeren (2004) develops a software component architecture dedicated to supply chain coordination, which supports distributed and cooperative organizations. Nevertheless, none of these authors consider cooperative and synchronization behaviors.

Finally, the third area concerns the Multi-Agent Systems (MAS) that describe coordination of supply chains. The MAS are regarded as one of the most promising technologies in supply chain management. Fox *et al.* (2000) model such a system with functional agents, which are responsible of several activities such as order acquisition, logistics, transport, or scheduling. The specificity of the agents is to support distributed decision making. A review of agent-based approaches in the supply chain management can be found in a paper of Parunak (1999) and a critical analysis in Caridi and

Cavalieri (2004). Some works are dedicated to specific supply chains. As an example, Moyaux and D'amours (2003) use a distributed application of the MAS to reduce the bullwhip effect in a forest supply chain. However, in those works, the distribution only concerns the decision making and does not concern the information network which is supported by a centralized architecture (Ahn and Lee, 2004).

In our approach, MAS is used in order to coordinate a multi-site supply chain with a distributed decision and distributed information architecture.

## 3. Simple industrial context: bronze tap production

### 3.1. Sample organization

A bronze tap production system is used to illustrate the problems of supply chain inside a complex network of enterprises. To do so, we consider the main components (SC) of a shipped tap (called PF in Figure 1):

- 1 Bronze body (called SC BBA)
- 2 O rings (called SC BA)
- 1 Blister (called SC A)

Those components are respectively made of:

- Base materials (copper and tin) (called SC BBAA)
- Rubber (called SC BAA)
- Cardboard (called SC AA)

These three elements appear in the product breakdown structure (PBS) or bills of material of the bronze tap.

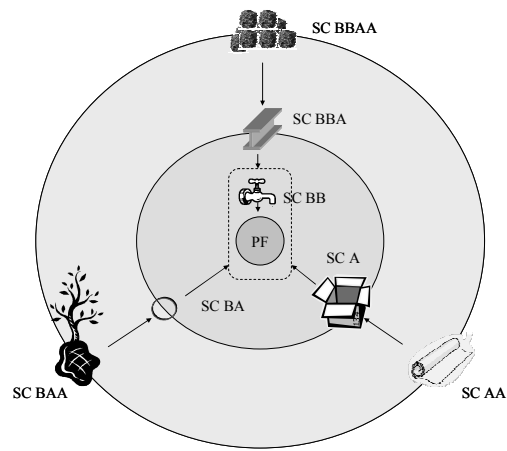

*Figure 1. Sold tap components*

For each element of the PBS corresponds a supplier entity, which can be either a single enterprise or a cluster of similar companies. From the knowledge of the product breakdown structure and the associated list of supplying firms, it is easy to model the corresponding tap's supply chain. The supply chain is then viewed as a set of successive tiers, in which each partner company is in relation with customers and suppliers on adjacent tiers. The adjacencies of tiers ($T_i$) for the supply chain of the bronze tap production are summarized in Figure 2.

In the reminder of the paper the enterprise network made by tiers one and two is considered (surrounded by the dashed line in Figure 2). This network is in relation with the external environment which is represented by direct customers (tier zero) and direct suppliers (tier three).

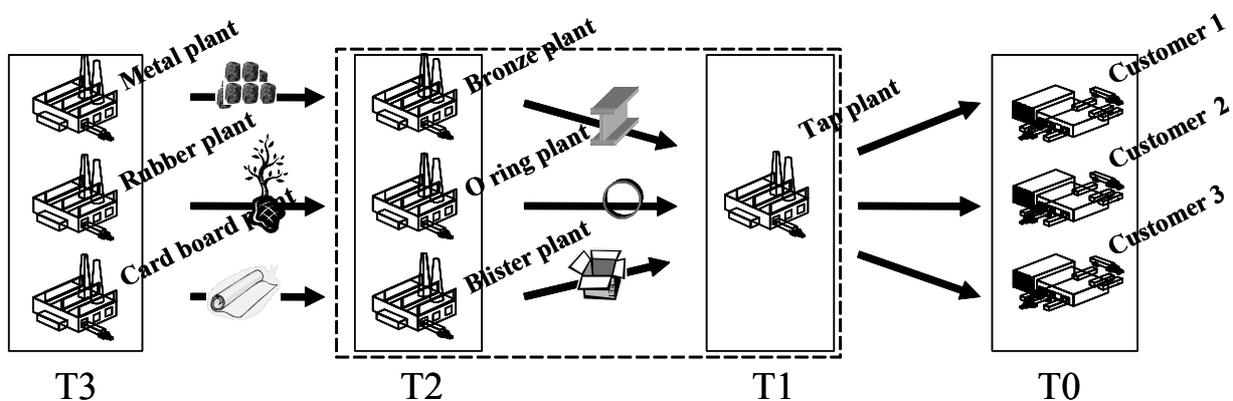

*Figure 2. Supply chain architecture*

## 3.2. The internal production processes

From the architecture presented in Figure 2, different actors are categorized and modeled, according to the owner internal process, *i.e.*, the tier position in the network architecture.

### 3.2.1. The tier 1 model of the tap plant

The tap production is divided in three activities:

- Body making;
- Assembly; and
- Finishing.

The assembly is the most important activity and corresponds more specifically to the bottleneck process.

The aim of the model is to elucidate if, from the current state of the system, a company β can accept or not a new demand from its client α. The decision is based on the evaluation of the work load on a production center. To rapidly determine in which conditions the company is able to manufacture the new order, a comparison is made between the added load induced by this new manufacturing demand and the idle state (*i.e.* unused production capacity) of each planning period. This analysis is done by focusing on the bottleneck activity of the internal production system.

Figure 3 shows the Petri net model of the production system of the tap plant as well as the propagation of the resource needs. The class Petri net used is Timed Place Object Petri Net (TPOPN). Each token has attached to it a data value representing time. So several time values can be used for one place (Monteiro and Ladet, 2001a).

The production process is made of sequences of stocks and activities. Each activity uses raw materials, from upstream stocks, and supplies downstream stock(s). The bottleneck activity uses SC BB and SC BA stocks to fill the SC B stock at the end of the chain.

When a client wants a quantity $Q$ of products PF at a date $Dd$, the tap plant requires stocks in SC A, SC BA and SC BBA for a quantity $Q3$ at $D3$, $Q2$ at $D2$ and $Q1$ at $D1$ respectively. The *Customer* and *Supplier* lines illustrate the external data exchanges of the tap plant with the tier 0 and tier 2.

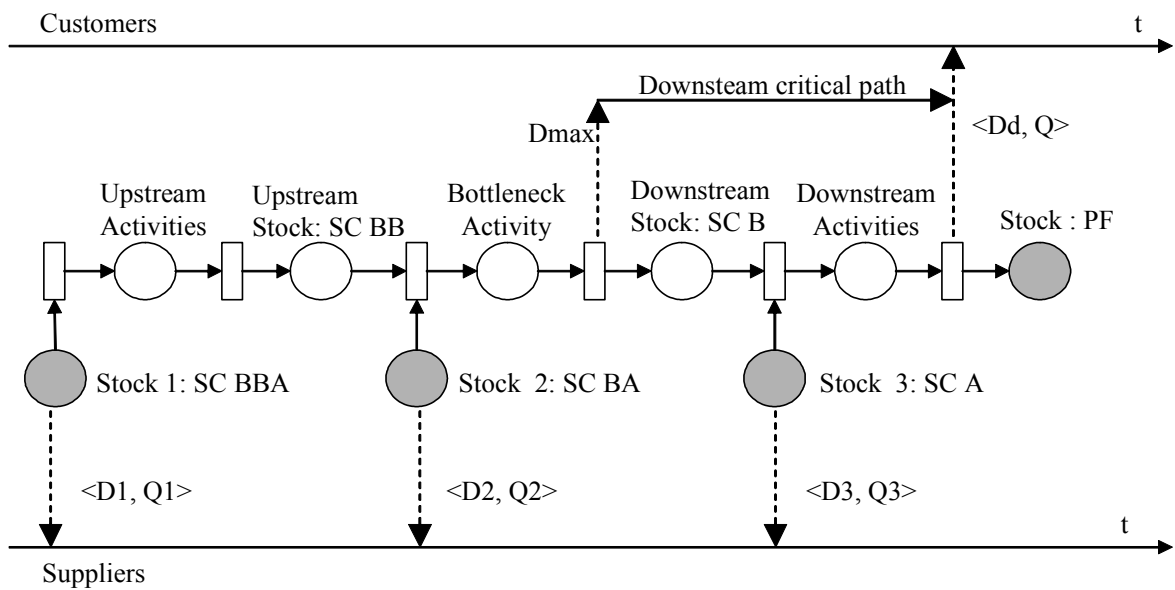

*Figure 3. Petri Net model of the Tier 1*

### 3.2.2. The tier 2 model of component supplier model

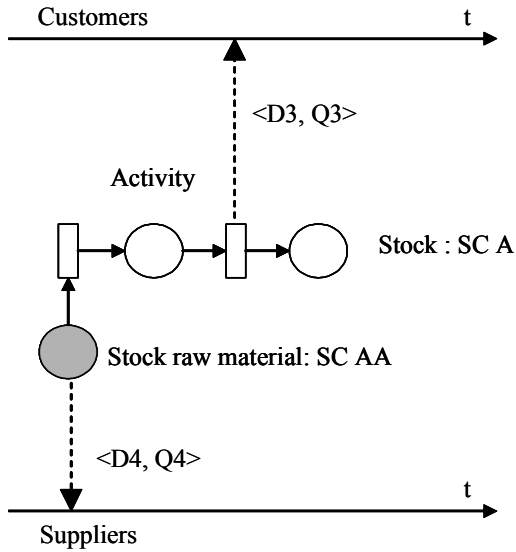

*Figure 4. Tier 2 model*

The production processes of the tier 2 suppliers are, a priori, distinct and complex. Thus, they can be modeled the same way as for the tier 1 supplier.

To simplify the illustration of this step and to illustrate the interdependence of the activities distributed through a logistic process, it has been decided to model the production systems with a demand propagation mechanism using temporal shift.

Figure 4 illustrates the internal process of a blister plant, which is modeled by a single activity. When one of the clients, for instance the tap plant, needs to be supplied at the date $D3$ with $Q3$ products of SC A, the blister plant has to be supplied itself at the date $D4$ with $Q4$ products of SC AA. This new demand is sent to the Tier 3 partner: cardboard plant.

### 3.3. Architecture definition

The supply chain is represented as a set of tiers (according to the product breakdown structure), in which each partner, defined as a Virtual Enterprise Node (VEN), is in relation with customers and suppliers on the adjacent tiers. It is assumed that each VEN is in relationship only with its adjacent VENs (*i.e.* no loop exists between the VENs) (OUZIZI, *et al.*, 2003). As a result, each VEN belongs to one tier. Figure 5a illustrates such a supply chain architecture described by VENs.

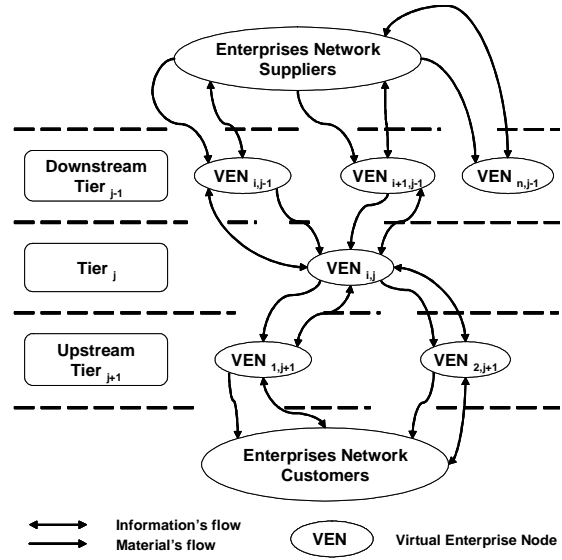

*a. Generic enterprise network architecture*

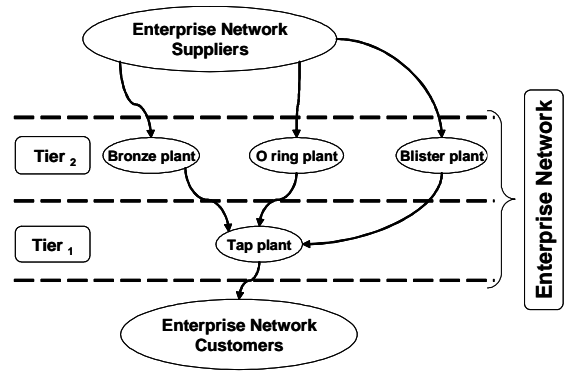

*b. Example of architecture*

*Figure 5. Enterprise Network*

Hardwick and Bolton (1997) introduce the concept of Virtual Enterprise (VE) to emphasize the idea of extended enterprise versus a centralized organization. The VE is different from the extended enterprise, which is organized around a central decision center, in the sense that it is an independent consortium that links its production means to increase its reactivity regarding to cope with its unpredictable environment.

Figure 5b illustrates the considered industrial example architecture following the generic architectural principles described above.

## 4. Modeling the agents

The enterprise network is modeled as a multi-agent system (Figure 6a), in which the agents use cooperative negotiation to establish a global consistent planning.

If a planning problem is detected, an agent called Tier Negotiator Agent (TNA) is activated. The purpose of this agent is to limit the negotiation process in terms of iterations and to facilitate cooperation between the VEN agents.

If the TNA cannot solve the problem, a Supply Chain Mediator Agent (SCMA) involving the whole enterprise network is used.

Furthermore, each VEN is made of:

- A Planner Agent (PA), used to establish the VEN planning with different parameters (costs, penalties, etc.). Depending on the enterprise planning strategy, different tools for planning could be used.

- A Negotiator Agent (NA), that receives orders, modifications of product demands and component deliveries from partner agents. It transmits requests to the planner agent and receives responses from it. It also transmits the results to the different partners. In case of a local problem, this agent contacts its TNA to obtain some help.

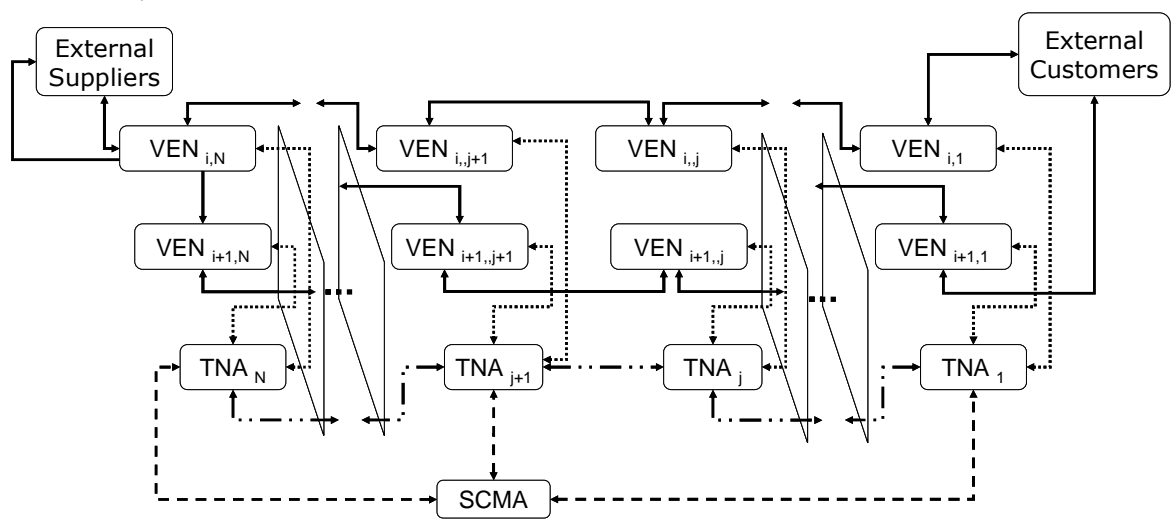

*a. generic agent architecture*

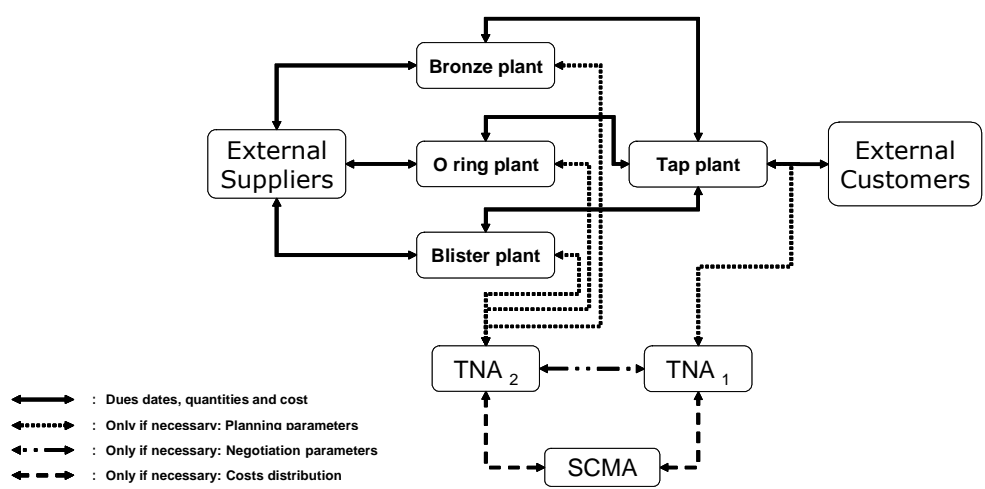

*b. example agent architecture*

*Figure 6. Agent Architecture representation*

Figure 6b illustrates the multi-agent system used in the industrial example architecture. Statecharts diagrams (Harel, 1986) are used to clearly illustrate the different functionalities of the agents. The statecharts focus on states and transitions of the system, and model the states and the complete flow of events.

### 4.1. The VEN

In principle, each VEN is facing internal and external constraints. The internal constraints are related to the capacity limits, whereas the external constraintsare related to:

- its customer VENs which require products in a minimal delay or with low costs for instance,
- its supplier VENs, which also have constraints of lead times, costs…

The VEN can be in two different situations depending on its capability to accept or to refuse the request. In the first situation, no consistency problem occurs. The VEN is used to propagate the client needs to supplier requests. In the second situation, a local problem occurs and a negotiation process has to be initiated.

Detecting consistency problems is the role of the Planner Agent, whereas local conflict management falls into the Negotiator Agent role.

Figure 7 shows the agent interactions. The VEN is limited by the rectangle area, which represents:

- the Negotiator Agent (NA),
- the Planner Agent (PA), and
- the internal information flows which include scenarios and requests.

Agents make decision on the basis of internal constraints, which depend on the production process and management for the PA and on strategic assessment for the NA.

The PA and NA agents use external tools to support planning and negotiation tasks. These agents are mainly generic ones, though using external tools allows them to be more adapted to face specific needs and practices of the enterprise. For instance, planning tools described by mathematical models (Dudek and Stadler, 2005) may be used in this approach. These tools may also be provided from the enterprise ERP system.

It is important to mention that only the NA is related to the external environment, and that the external environment is made of VENs of direct partners and a Tier Negotiator Agent (TNA).

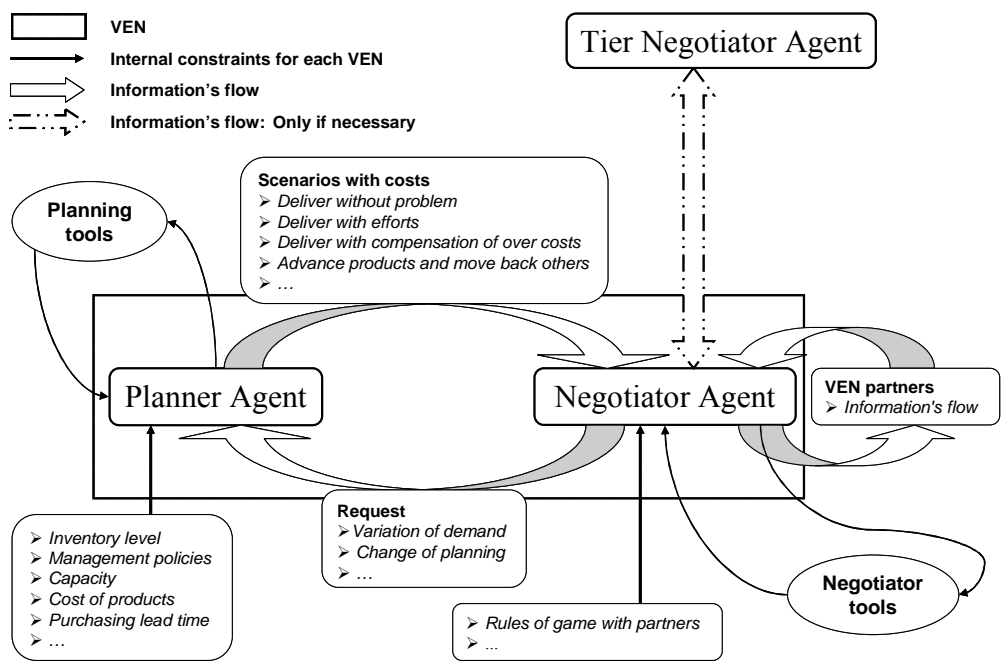

*Figure 7. Planning and negotiation processes of VEN*

### 4.1.1. The Negotiator Agent (NA)

Statechart modeling is used to depict the negotiation process happening within the VEN. Figure 8 shows different messages used by the NA, which correspond to different states of negotiation with downstream or upstream tiers and its TNA.

Six types of external messages can modify the NA state:

- Order messages of type C_US. In thit case, the client sends product requirements to the NA. This order can be a new one or be a modification to an existing one.

- Answer messages to a product modification of type RN_US. The client sends its reply about a modification request (N_US message).

- Messages of type R_PA. This type of messages is sent by the PA as a response to D_PA_N, D_PA_M or D_PA_A (see 4.1.2). Such a message can be a response to an Upstream (_US) or Downstream (_DS) request. In the case of R_PA_US(n), the message suggests different scenarios, where each scenario specifies the requirements curves as input (demand quantities per time period), the proposed deliveries as output, the production costs, and the over costs generated in terms of the overtime, subcontracted quantities…

- Downstream modification messages of type N_DS. The supplier sends product delivery modification to the NA as a set of potential scenarios.

- Messages of the type D_TNA. The message is send by the TNA to the NA. In case of problem somewhere in the tier, the TNA needs to know the actual state of each VEN production. The D_TNA message allows the TNA to retrieve that information.

- Messages of type C_TNA. When a TNA has solved a problem, it sends to each VEN an order according to the assumed solution.

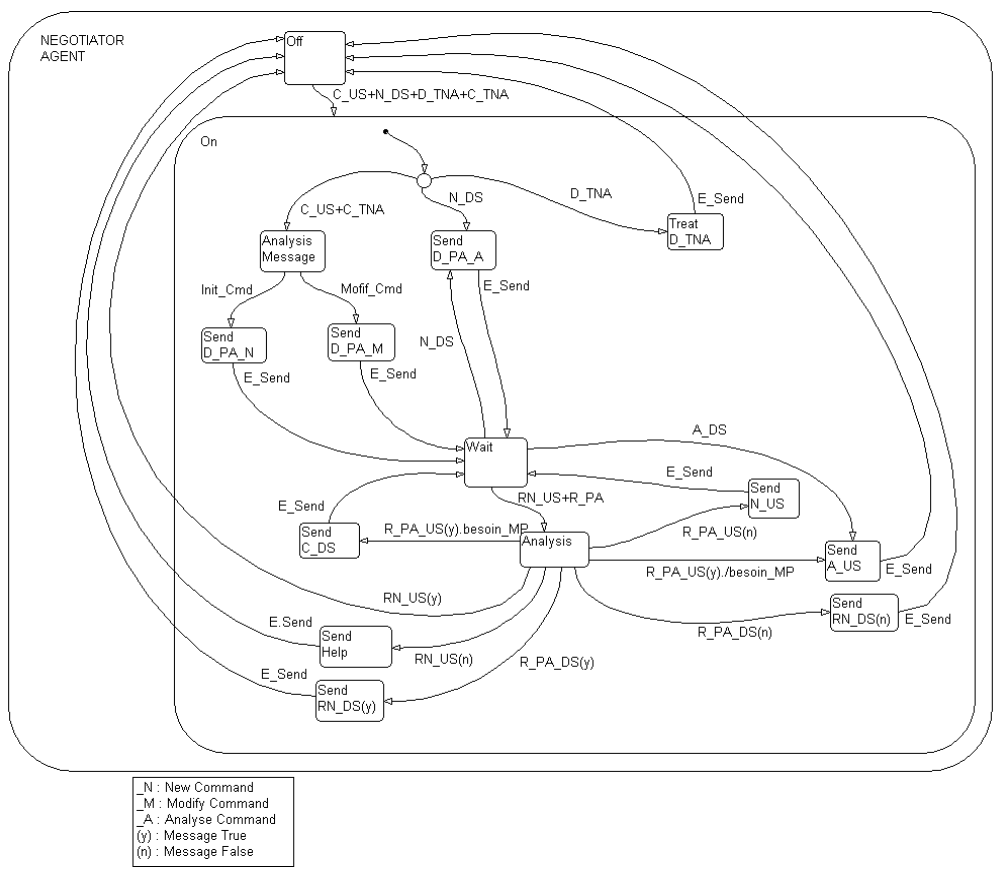

*Figure 8. NA statechart*

### 4.1.2. The Planner Agent (PA)

The PA role is to determine if any modification can be supported or not by the production process. This internal agent is initiated only by VEN NA. The PA makes the interface between the internal and the external environments. The NA brings external constraints in terms of delays, cost, etc. The PA has also to take into account internal constraints in terms of production load, capacity…

Three types of external messages issued from NEV NA can modify the PA state:

- New order messages of type D_PA_N: Once the VEN NA receives a message of C_US (or C_TNA) type (see 4.1.1), which concerns a new order, the VEN NA forwards it to the PA to specify the production needs and the conditions (such as the limit of overtime, quantities subcontracted, delay penalties …).

- Upstream modification messages of type D_PA_M. Once the VEN NA receives a message of C_US (or C_TNA) type (see 4.1.1), which concerns an existing order modification (see 4.1.1), it sends a D_PA_M message to the PA to specify the production modifications.

- Downstream modification messages of type D_PA_A. Once the VEN NA receives a N_DS type message (see 4.1.1), it sends a D_PA_A message to the PA to specify production scenarios.

### 4.2. The Tier Negotiator Agent (TNA)

When a $VEN_{ij}$ (Figure 6a) of a considered tier is unable alone to find a valid planning, it forwards the problem to its $TNA_j$. The TNA has a total view on its level. Thus, the TNA can be informed, by questioning its VEN, of the system state and the local blocking causes.

The TNA first objective is to solve the problem at its own tier level. For instance, as a first goal, the $TNA_j$ proposes a load distribution for several VENs in order to able to carry out the blocking production situation. If no solution is found, the goal becomes to initiate and manage negotiations with its direct TNA partners. To do so, $TNA_j$ is linked with the $TNA_{j+1}$ and the $TNA_{j-1}$. This set of negotiations allows recursive propagation of the problem until its resolution. If the $TNA_j$ cannot solve the problem, the Mediator Agent (MA), involving the whole enterprise network, is requested.

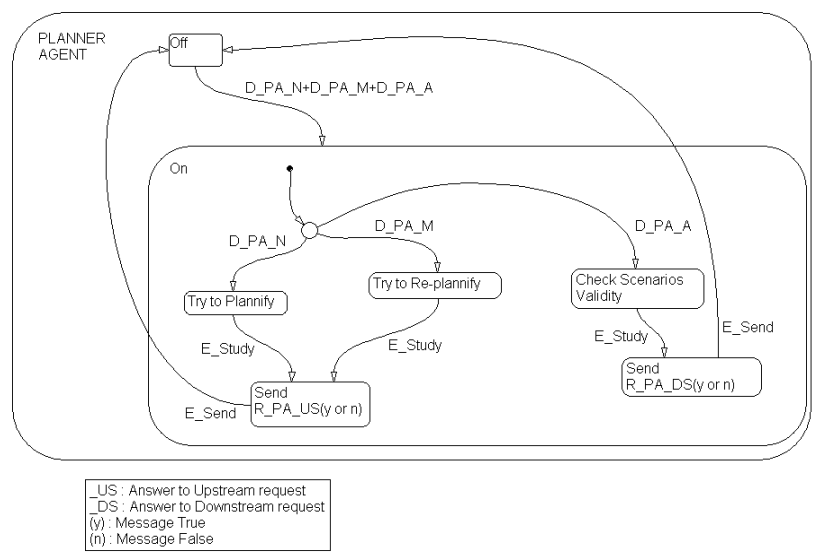

*Figure 9. PA statechart*

### 4.3. The Supply-Chain Mediator Agent (SCMA)

The goal of the SCMA is to solve conflicts by relaxing constraints, relaxation is based on a global cost.

The main objective of an enterprise network is to minimize the purchase and the production costs as well as to ensure a positive benefit (Anciaux, *et al.*, 2003).

The global benefit of the network is then a function of the total selling and the total cost, such as:

$$\sum_{all\ VENs} selling - \sum_{all\ VENs} costs \geq 0 \qquad (1)$$

While preserving benefit, the SCMA authorizes local deficit on one or more VENs which are responsible for the blocking situation. The SCMA distributes this deficit among all the partners and manages penalties which are induced by not respecting the partner contracts. In our approach, to solve this it has been chosen to apply the "win-win" principle on the long run. This principle prevents to always penalize the same company and allows preserving advantages in a durable matter relation.

### 5. Illustrations

To illustrate the behavior of the different agents in a multi-agent architecture, let us consider a request for a new product. This illustration is described using UML activity diagrams. Three business entities are considered: the VEN Client, the VEN Tap Plant and the VEN Blister Plant (see Figure 10).

The UML activity diagrams have been established from the Tap Plant point of view. This is why there is only one VEN where both PA and NA appear. In the swim lanes of the other VEN, activities represented by black boxes, are not detailed. Effectively, those activities are internal processes of the Blister Plant and Client and are not involved in the Tap Plant environment. Moreover, the UML stop symbol illustrates the end of negotiation for the considered entity.

In these illustrations, the Tap Plant NA receives a new order from the client NA (message #1 in Figure 10 and Table 1). The Tap Plant PA, using planning tools (see Figure 7), can determine if this new induced load respects internal constraints. In this example, the new order can be planned, but an external product, such as blister (SCA), is required. Consequently, a request is sent to the Blister Plant NA (message #4). Several answers can follow (messages 5 or 5') as illustrated in the following cases.

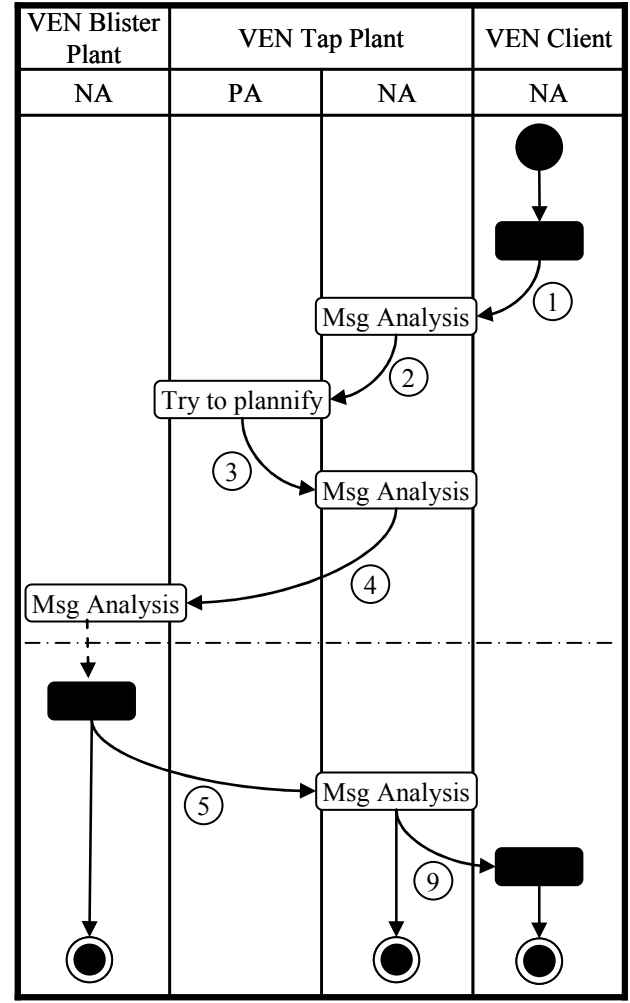

*Figure 10. Swimlanes in activity diagram for a new product request in a normal mode*

Three main situations which can occur after a request for a new product can be described:

- Normal mode,
- Delivery problem with scenario acceptation
- Delivery problem without scenario acceptation

Table 1 details four transition messages common in the three main situations.

| N° | Message / Comment |
|---|---|
| 1 | C_US(PF,<06/01,100>): Demand of 100 units of PF for the 1st of June. |
| 2 | D_PA_N(PF,<06/01,80>): Demand to plan the production of 80 PF (20 PF already in stock). |
| 3 | R_PA_US(y).(SCA<05/28,60>): Able to produce if 60 units of blister (SCA) available for the 28th of May. |
| 4 | C_DS(SCA,<05/28,60>): Demand of 60 units of SCA to Blister Plant NA. |

*Table 1: Flow definition*

### 5.1. Normal mode

In the normal mode, illustrated by Figure 10, the Blister Plant answer allows to satisfy Tap Plant and Client needs. In this case, a Client new order is possible without modifications and an agreement message is sent to it (message #9 in Table 2). Issuing this message ends the negotiation process.

| N° | Message / Comment |
|---|---|
| 5 | A_DS(SCA,<05/28,60>): The external product (SCA) could be delivered at the due date with the desired quantity. |
| 9 | A_US(PF,<06/01,100>): The Client PF request is feasible and a contract is concluded. |

*Table 2: Flow definition for case #1*

Now, the Blister Plant can not satisfy the initial needs of the Tap Plant and eventually of the client. In such a situation, the Blister Plant NA sends a proposal (message #5', see Figure 11 and Figure 12) which is analyzed by the Tap Plant PA.

### 5.2. Delivery problem with acceptation scenario

The message #5' (described in Table 3) contains several scenarios which constitute a number of different proposals. Each scenario is composed of a delivery date and a corresponding product quantity.

If the Tap Plant PA analysis concludes that one of the proposed scenarii is acceptable, an agreement is sent to the Blister Plant (message #8) and to the Client (message #9).

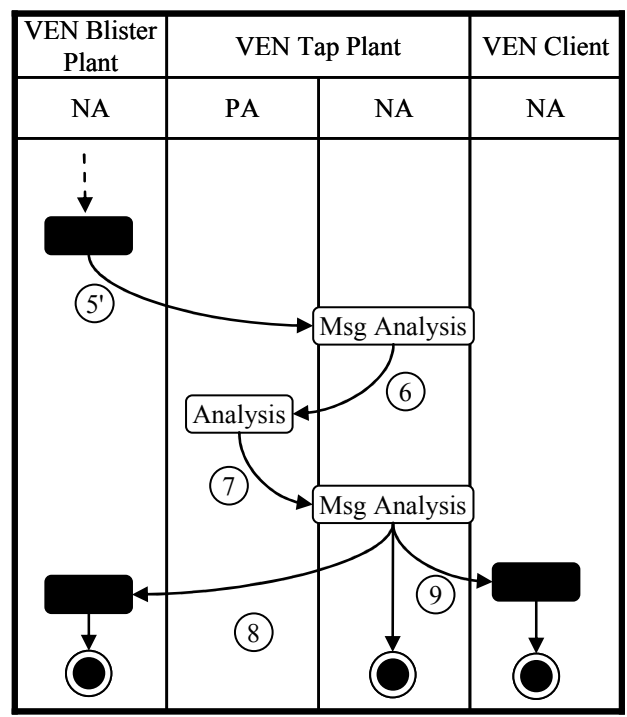

*Figure 11. Swimlanes in activity diagram for product requirement with scenario acceptation*

| N° | Message / Comment |
|---|---|
| 5' | N_DS(SCA,(<05/28,40>,<05/30,20>)(<05/30,50>)): The external product (SCA) request could not satisfied. Two scenarios are proposed instead: First, only 40 units of SCA should be delivered at the due date, the rest (20) should be available only two days after. Second, the total demanded quantity should be delivered the 30th of May instead of the 28th. |
| 6 | D_PA_A(SCA,<05/28,40>,<05/30,20>(<05/30,60>)): The Tap plant NA transfers the scenarii to its PA to check if at least one of them is acceptable. |
| 7 | R_PA_DS(y,<05/28,40>,<05/30,20>): The 1st scenario is acceptable (*i.e.* Tap Plant will be able to respect its own constraints about PF request, despite of SCA delivery delay). |
| 8 | RN_DS(y,<05/28,40>,<05/30,20>): The Tap Plant accepts the Blister conditions. A contract is concluded. |

*Table 3: Flow definition for case #2*

## 5.3. Delivery problem without acceptation scenario

In this last example, the Tap Plant PA analysis concludes that none of the proposed scenarii is acceptable. The Tap Plant PA then notifies the Blister Plant PA of the refusal (message #8'). At this point, the system is in a local blocking situation. In order to relax the constraint, the Blister Plant NA sends a request to its TNA, which tries to solve the problem as explained in § 4.2. Figure 12 illustrates this case, where the Client is not shown, since it is not involved, but the TNA entity is.

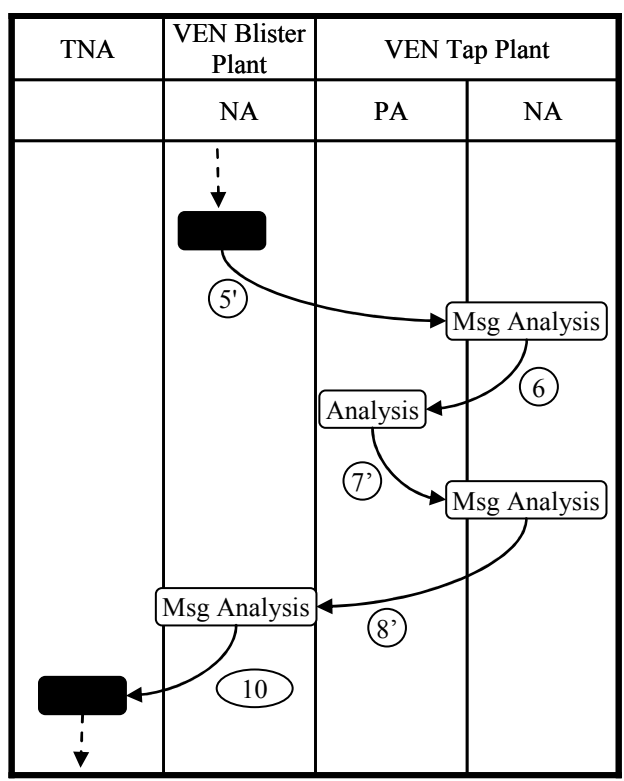

*Figure 12. Swimlanes in activity diagram for product requirement with TNA use*

| N° | Message / Comment |
|---|---|
| 7' | R_PA_DS(n) : Any scenario is acceptable. |
| 8' | RN_DS(n): Tap Plant informs the Blister Plant its refusal. |
| 10 | Send "help" to TNA. |

*Table 4: Flow definition for case #3*

## 6. Conclusion and future research

New enterprise network organizations, based on cooperation, face problems of flow control and management because of independent decision centers. To improve the productivity and the reactivity of these networks, decision distribution along the supply chain needs to be clear and coherent between the partners.

A new approach has been developed that manages multi site resource planning to coordinate the needs of the network. The proposed architecture is based on a Multi Agent System (MAS) paradigm. On standard running, the MAS guarantees the intrinsic decision distribution among each partner. If a problem occurs locally, the MAS is able to find a global solution.

A Virtual Enterprise Node (VEN) is defined as an individual actor of the network. A VEN is either a Planner Agent that allows a dynamic planning based on local constraints (production cost, load, capacity ...) or a Negotiator Agent that is in charge of the cooperation with the partners. In addition to the VEN agents, two virtual agents complete the architecture. At a higher level of the VEN, the Tier Negotiator Agents are in charge of relaxing the constraints at tier level. At the uppermost level, the Supply Chain Mediator Agent has to find a global solution via cost based on constraint relaxation. As a result, these actors integrate a hierarchical architecture that guarantees robustness and flexibility in the structure of the enterprise network.

Some current evolutions of this system strictly concern the network of suppliers, but could be extended to the network of distributors. Some particular architecture elements, like skill centers, need to be studied. In addition, in our approach the coordination mechanisms are based on scenario exchanges. Another future research could look at other coordination mechanisms, such as the simplest mechanism of single acceptance and refusal exchange. More sophisticated mechanisms can also be taken into account, like the coordinate planning proposed by Dudek and Stadler (2005) in which a more integrated relationship is represented. All these coordination mechanisms have intrinsic performances and need to be analyzed and compared with our approach.

Another extension may concern the use of web services over an http network instead of MAS. Nevertheless, this extension needs an analysis of data flow reliability in such networks.